\documentclass[twocolumn, prl, showpacs, superscriptaddress, preprintnumbers, floatfix, amsmath,amssymb]{revtex4}                     

\usepackage[dvips]{graphicx}                   
\usepackage{epsfig}
\usepackage{dcolumn}                           
\usepackage{bm}                                
\usepackage{setspace}
\usepackage{comment}
\usepackage{color}
\definecolor{mycolor}{rgb}{0.1,0.1,0.5}
\definecolor{mycolor}{rgb}{0.0,0.0,0.0}       
\usepackage[colorlinks=true, citecolor=mycolor, linkcolor=mycolor, urlcolor=mycolor, hyperindex=true]{hyperref}
\usepackage{natbib}

\begin{document}

\title{Magnetic domain fluctuations in an antiferromagnetic film observed with coherent resonant soft x-ray scattering}

\date{\today}

\pacs{75.10.Nr, 75.70.Rf, 78.70.Ck}                

\author{S. Konings}
\affiliation{{Van der Waals-Zeeman Institute, University of Amsterdam, 1018 XE Amsterdam, The Netherlands}}
\author{C. Sch\"{u}{\ss}ler-Langeheine}
\affiliation{{II. Physikalisches Institut, Universit\"{a}t zu K\"{o}ln, Z\"ulpicher Str. 77, 50937 K\"{o}ln, Germany}}
\author{H. Ott}
\affiliation{{II. Physikalisches Institut, Universit\"{a}t zu K\"{o}ln, Z\"ulpicher Str. 77, 50937 K\"{o}ln, Germany}}
\author{E. Weschke}
\altaffiliation[Present address: ]{Helmholtz-Zentrum Berlin f\"ur Materialien und Energie GmbH}
\affiliation{{Institut f\"{u}r Experimentalphysik, Freie Universit\"{a}t Berlin, 14195 Berlin, Germany}}
\author{E. Schierle}
\altaffiliation[Present address: ]{Helmholtz-Zentrum Berlin f\"ur Materialien und Energie GmbH}
\affiliation{{Institut f\"{u}r Experimentalphysik, Freie Universit\"{a}t Berlin, 14195 Berlin, Germany}}
\author{H. Zabel}
\affiliation{{Lehrstuhl f\"ur Experimentalphysik/Festk\"orperphysik, Ruhr-Universit\"at Bochum, 44780 Bochum, Germany}}
\author{J. B. Goedkoop}
\affiliation{{Van der Waals-Zeeman Institute, University of Amsterdam, 1018 XE Amsterdam, The Netherlands}}

\begin{abstract}
We report the direct observation of slow fluctuations of helical antiferromagnetic domains in an ultra-thin holmium film using coherent resonant magnetic x-ray scattering. We observe a gradual increase of the fluctuations in the speckle pattern with increasing temperature, while at the same time a static contribution to the speckle pattern remains. This finding indicates that domain-wall fluctuations occur over a large range of time scales. We ascribe this non-ergodic behavior to the strong dependence of the fluctuation rate on the local thickness of the film.
\end{abstract}
\maketitle
\newpage

Slow dynamics of magnetic domains on time scales of nanoseconds
and longer are of high practical importance. Domain wall dynamics
play a crucial role in magnetization reversal processes; thermally
activated domain wall motions determine the lifetime of
magnetically stored information. Slow dynamics on nanometer length
scales is best probed by x-ray photon correlation spectroscopy
(XPCS) \cite{dierker:95a, thurnalbrecht:96a, mochrie:97a,
malik:98a,seydel:01a,sikharulidze:02a,sikharulidze:03a,
madsen:04a,madsen:05a,shpyrko:07a,turner:08a} using the coherent diffraction
or \textit{speckle} pattern, which is generated when a coherent
light beam scatters from a disordered structure. Any fluctuations
in the disorder lead to a change in the speckle pattern; the
dynamics in the sample can be obtained by measuring the time
averaged intensity correlation function
(ICF) \cite{berne:00a,pusey:89a, kroon:96a} of the speckle
intensities on time scales ranging from 50 ns
\cite{sikharulidze:02a}  to hours \cite{malik:98a,shpyrko:07a}.
Importantly, with PCS one can obtain directly the fluctuating and
static parts of the sample \cite{pusey:89a,kroon:96a}, which makes
PCS highly attractive for the investigation of systems where
pinning or jamming effects occur \cite{malik:98a}.

In order to study magnetism the speckle experiment has to be
sensitive to spin degrees of freedom. X-ray scattering in the
conventional x-ray range, even at electronic resonances, has a low
magnetic scattering cross section \cite{vettier:01a}. The only
exception are $5f$ systems where a magnetic phase transition in
UAs has indeed been observed by the loss of speckle contrast
\cite{yakhou:01a}. But actinide systems are only of limited
practical relevance, whereas in most interesting magnetic systems
the magnetism is carried by $3d$ or $4f$ electrons. $4f$ magnetism
can be probed at the $2p \rightarrow 5d$ resonances in the
conventional x-ray range, but such experiments usually require
polarization analysis of the scattered photons, which is hard to
combine with the high spatial resolution required to resolve the
speckle pattern. The scattering cross section for $3d$ magnetism
on and off the $1s \rightarrow 4p$ resonance is very low and only
few magnetic scattering studies of $3d$ transition-metal systems
in the conventional x-ray range exist.

In some cases one may probe magnetism indirectly, via its coupling
to structural degrees of freedom \cite{shpyrko:07a}. But while a
coupling of spin order to a charge density wave or to charge order is found for
many systems, both orders will generally form on different
temperature scales. Examples are layered nickelates
\cite{yoshizawa:00a} or cobaltates \cite{zaliznyak:00a}, for which
the temperature scales for charge and for spin order are clearly
different. An extreme case is La$_{1.5}$Sr$_{0.5}$CoO$_4$, where
charge order sets in below 750 K, while static spin order is not
observed above 35 K \cite{zaliznyak:00a}. It is therefore not to
be expected that the charge dynamics are generally representative
for the dynamics of magnetic order. This means that the indirect
approach cannot be generally applied and one needs to probe the
magnetic signal directly.

This is possible in soft x-ray range where resonant scattering provide a high
magnetic contrast. PCS at the Co $2p \rightarrow 3d$ resonance was used to
study the influence of disorder on the static domain pattern of Co/Pt
multilayers for different magnetic fields \cite{pierce:05a}. In this Letter we
show that using soft x-ray PCS it is actually possible to directly probe
\emph{fluctuating} magnetic domains near a second-order phase transition and to
address the question, how this transition is affected by static disorder in the
system.

We studied an 11 monolayers (ML) thin epitaxially grown Ho-metal film
sandwiched between Y-metal layers \cite{ott:06a,leiner:00a}. The structural film properties were characterized with x-ray reflectivity measurements in the conventional x-ray range. We found that the roughness at the two Ho/Y interfaces causes thickness variations over the film of 2 monolayers. The Ho film grew pseudomorphically on Y \cite{bentall:03a} with a structural in-plane correlation length of about 80 nm, which is only slightly shorter than that of the Y layers (100 nm). 
\begin{figure}
\centering \includegraphics[width=0.42\textwidth]{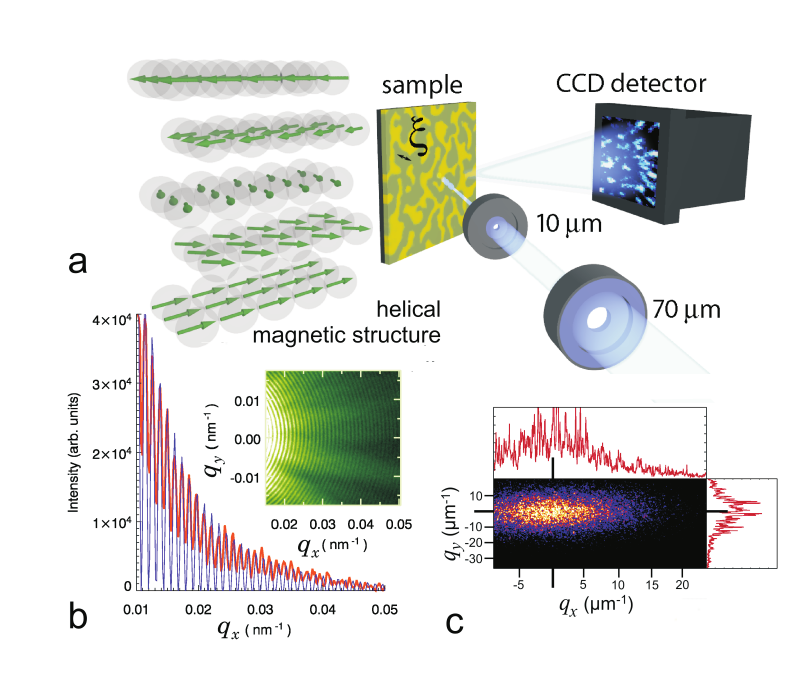} \caption{($a$) Sketch of the
scattering experiment and the helical antiferromagnetic structure. Two pinholes
select the spatially coherent part of the undulator radiation. The correlation
length $\xi$ in the sample is indicated with a black line and is in reality
$\sim$100 times smaller than the beam diameter.
\\
($b$) The diffraction pattern of the 10~$\mu$m pinhole. $q$ denotes the
momentum transfer.  A line cut through the center is shown in red, together
with a least squares fit of an Airy pattern in blue. From the contrast we determine the coherent fraction of the light to be 40 \%.
\\
($c$) A typical speckle pattern of the magnetic satellite peak at 30~K. The
two intensity line traces plotted in red go along $q_x$ and $q_y$ through the center of the peak. The tick marks for $q_x$ refer to $q_y = 0$ and vice versa.} \label{fig:fig1}
\end{figure}
Holmium metal
displays a helical magnetic phase (sketched in Fig.~\ref{fig:fig1}(a)) over a
wide temperature range leading to superstructure peaks in the magnetic
diffraction signal separated by a wavevector $(0,0,\pm \epsilon)$ from the
structural peaks \cite{helgesen:94a}. The ordering temperature $T_N$ depends on
the film thickness such that films below 10 ML, which is about one helix period
length in bulk Ho, do not show any helical order \cite{weschke:04a,cinti:08a,cinti:09a,cinti:09b}. The 11-ML film studied here
is hence near the stability limit for helical order and thus close to
two-dimensionality. On the other hand the stability of the helical phase in such a film should be very susceptible to
slight thickness variations of the Ho layer because $T_N$ is a steep function of the thickness.

The experiments were carried out at the U49/2-PGM1 and UE46 beam lines
of BESSY (Helmholtz Zentrum Berlin) using the soft x-ray diffractometer built at the FU
Berlin. In order to observe the magnetic signal from the film, we
used the strong magnetic contrast that is found at the \mbox{$3d
\rightarrow 4f$} ($M_5$) excitation in the soft x-ray range at a
photon energy around 1344~eV corresponding to a photon wavelength
$\lambda$=9.2~{\AA} \cite{ott:06a, konings:07a}.

\begin{figure}
\includegraphics[width=0.35\textwidth]{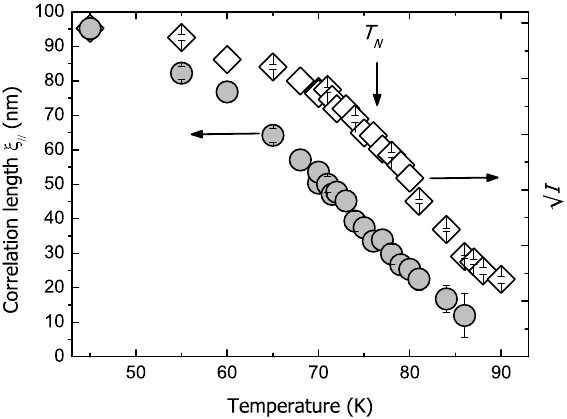}
\caption{In-plane magnetic correlation length $\xi_{||}$
(circles and left axis)) and square root of the integrated
satellite intensity (diamonds and right axis) as measured
with a non-coherent beam.} \label{fig:fig2}
\end{figure}

We intercepted the first-order $(00\epsilon)$ satellite with a direct-exposure
soft x-ray CCD camera as displayed in Fig.~\ref{fig:fig1}(a). With incoherent
light, we observe a smooth diffraction peak. As is shown in Fig.~\ref{fig:fig2}
(diamond symbols), the scattered intensity in this peak decreases when the
nominal transition temperature $T_{N}$=76~K is approached, but does not vanish
up to $\approx$ 90 K \cite{ott:04a}. The peak profile is well described by a
single Lorentzian, with a half-width that equals the inverse of the
in-plane correlation length $\xi_{||}$ of the magnetically ordered regions. As shown
by the circular symbols in Fig.~\ref{fig:fig2}, we find $\xi_{||}$ to decrease from
90~nm (similar to the structural correlation length) at 45~K to 10~nm at 85~K. The loss of correlation already sets in around 50
K and $\xi_{||}$ keeps on changing over an unusually wide temperature interval
of more than 40 K. This may either be an intrinsic effect caused by the
proximity of the film to two-dimensionality or an effect of static disorder
induced by the interface roughness.
In order to disentangle the roles of static and dynamic effects at
this second-order type phase transition, we selected the
transversely-coherent fraction of the BESSY~II undulator radiation
using a 10~$\mu$m pinhole in front of the sample \cite{coherence}.
This causes the smooth magnetic diffraction peak to break up into
a myriad of speckles [Fig.~\ref{fig:fig1}(c)], which form the
diffraction pattern of the magnetic domain structure of the
particular illuminated spot. Speckle fluctuations at different
distances from the peak center are related to real-space
fluctuations on different length scales [Fig.~1(c)]. We obtain the
most intense signal from magnetic disorder on length scales of
more than 100 nm, which we assign to helicity or phase domains.

\begin{figure}
\includegraphics[width=0.35\textwidth]{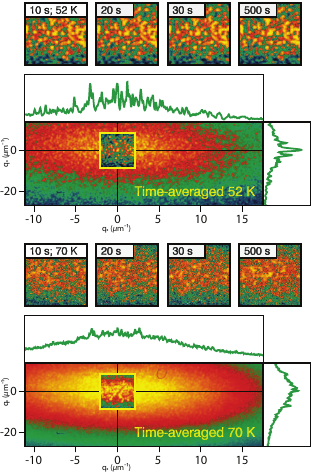}
\caption{Time-averaged
intensity  distribution of the magnetic satellite peak (\textit{large panels})
at 52 K (a) and 70 K (b). A logarithmic color scale is used to better observe
the speckle at higher $q_{||}$. The tick marks for $q_x$ refer to $q_y = 0$ and vice versa. The snapshots (small panels) are single frames
with an exposure time of 10 s taken at the indicated time. For the snapshots a
linear color scale is used.}
\label{fig:fig3}
\end{figure}

We followed the time evolution of the speckle pattern by recording movies over
a period  of several hours with exposure times of 4 or 10~seconds. Snapshots
from these movies for 52~K and 70~K are presented in the small frames of
Fig.~\ref{fig:fig3}. The complete movies for various temperatures are available
online \cite{web2}. At 52~K the speckle pattern is static on a time scale of
one hour. At intermediate temperatures the speckle pattern starts to change
with time and already at 70~K the movement of the speckles is very vivid.

In order to quantify how much of the speckle pattern is moving, we
took the time  average of all the frames in one movie, shown in
the large panels of Fig.~\ref{fig:fig3}. At 52~K the average
pattern is equal to that of a single frame. Closer to the phase
transition subsequent speckle patterns differ strongly, and the
time-averaged speckle pattern is much smoother than the individual
frames thus showing that domain-wall fluctuations have started.
But even the time averages of films at higher temperatures over
hours show some graininess due to the existence of static speckles
connected to non-fluctuating parts of the domain pattern. These
static speckles are found on all length scales, i.e. at all
distances from the peak center, which is very obvious in the line
cuts in Fig.~\ref{fig:fig3}(b). Our finding thus implies that some
regions of the sample are fluctuating, while others remain fixed
over the measurement period: the system behaves non-ergodic.

We argue that the most likely cause for the observed wide range of fluctuation times are variations in the Ho film
thickness. As noted above, the ordering temperature $T_{N}$ of Ho films is
critically dependent on the film thickness in the range of 10 to 12 monolayers
\cite{weschke:04a,cinti:08a,cinti:09a,cinti:09b}. This leads to a picture in which at low temperatures the magnetization has settled down in an irregular static domain structure with the magnetic domain size determined by the structural in-plane correlation length of the film. Domain walls exist between regions of opposite helicity or helical order phase slips  As the temperature increases, the thinnest regions approach their local N\'{e}el
temperature and start to fluctuate. At higher temperatures, gradually the
thicker regions join in, explaining the observed reduction of the magnetic
correlation length and the increasing fluctuation rates in the speckle movies.

\begin{figure}[!t]
\includegraphics[width=0.42\textwidth]{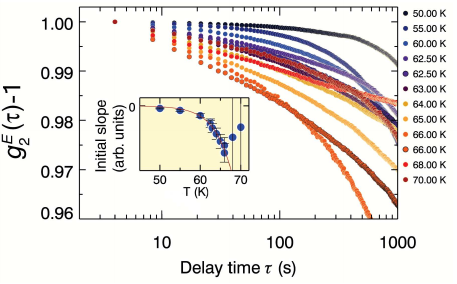}
\caption{Normalized
experimental results for \mbox{$g_{2}^{E}(\tau)-1$} at the
indicated temperature. The inset shows the initial slope for each
temperature as determined by a linear fit; error bars give the
quality of the fit.  The red line is an exponential fit to these
data points.} \label{fig:fig4}
\end{figure}

The increase of the fluctuation rates with temperature are also
reflected in the  time-averaged intensity correlation function
(ICF). We performed the ensemble averaging required for
non-ergodic systems \cite{pusey:89a,kroon:96a} by averaging over
all the pixels with a distance from the peak center of less than
0.004~nm$^{-1}$ corresponding to correlation lengths of 250 nm and
larger. The normalized ensemble averaged ICF is defined as
\mbox{$g_{2}^{E}(\tau)=\langle I(t)I(t+\tau)\rangle_{E} / \langle
I(t) \rangle_{E}^2 $}, where $\tau$ gives the delay time between
two data samples and the brackets $\langle \; \rangle_{E}$
indicate time and ensemble averaging. In Fig.~\ref{fig:fig4} we
show the results for \mbox{$g_{2}^{E}(\tau)-1$} for the different
temperatures, with the curves normalized to the first data point.
For long correlation times the curves show an unsystematic
behavior, which we ascribe to slow drifts of the beam line and
setup. For short correlation times, however, the signal changes
faster and faster with $\tau$ upon heating. This we quantified by
determining the slope of \mbox{$g_{2}^{E}(\tau)-1$} in the first
50 seconds as depicted in the inset. The change of slope reflects
an increase of the domain wall dynamics that speeds up with
increasing temperature. The last two points in the inset indicate
that for temperatures above 66~K the scattered intensity becomes
too low for a reliable analysis. \cite{lumma:00a, falus:06a}

In conclusion, we show that already with the limited coherent flux available at
present 3$^{rd}$-generation light sources, x-ray photon correlation
spectroscopy at soft x-ray  resonances provides unique information on the
influence of disorder on magnetic phase transitions. In general, such transitions are analysed in terms of scaling theory and critical exponents, which is only relevant for strictly ergodic systems. The non-ergodic behavior observed here in an ultrathin Holmium film is prototypical for systems in which domain walls are pinned in the potential landscape formed by disorder, in this case due to local thickness variations that cause local variations in $T_N$. This image also provides a natural explanation for the smearing out of the magnetic phase transition of the film over a wide temperature range found for this sample.

Finally, we wish to point out that resonant soft x-ray scattering is also sensitive to charge and orbital order. Soft x-ray resonant PCS therefore provides a new experimental approach for the study of the phase transitions that are found abundantly in complex correlated electron compounds. Many of these systems have chemical disorder due to doping, which provides an intrinsic pinning landscape. For non-ergodic systems a characterization of the fluctuation rate on a wide range of time and length scales is necessary. Clearly, with the new x-ray free-electron laser sources, which deliver a fully coherent photon beam \cite{grubel:07a}, the study of more highly interesting systems with soft x-ray PCS will become possible.

\begin{acknowledgments}
The authors thank the BESSY and HZB staff of U49/2-PGM1 and UE46-PGM,
notably D. Schmitz for expert assistance during the measurements. We thank G. Wegdam and D. Bonn for discussions on PCS. The authors gratefully acknowledge financial support by the Stichting voor Fundamenteel Onderzoek der Materie (FOM), the Deutsche Forschungsgemeinschaft (DFG) through SFB 608, the BMBF through projects 05ES3XBA/5, 05KS7PC1, 05KS7PK1, and BESSY as a partner of the EU's I3 IA-SFS project (BESSY-ID.06.2.198; \mbox{RII 3CT-2004-506008}).
\end{acknowledgments}


\begin{thebibliography}{28}
\bibitem[{coherence()}]{coherence}
\bibinfo{howpublished}{The coherent flux was $10^9$ ph/s, the bandwidth $E/{\Delta E} = 10^4$, the longitudinal coherence length was about 10 microns. The coherent fraction of the radiation as determined from the contrast in the Airy pattern was 0.4.}
\end{thebibliography}

\end{document}